\documentclass[journal]{IEEEtran}
\ifCLASSINFOpdf

\else

\fi
\usepackage{mathrsfs}

\usepackage{amsmath}
\usepackage{ntheorem}

\usepackage{makeidx}  
\usepackage{algorithm}
\usepackage{algorithmic}
\usepackage{graphicx}
\usepackage{subfigure}
\usepackage{epstopdf}
\usepackage{bm}
\usepackage{cite}
\usepackage{stfloats}

\usepackage{amssymb}
\usepackage{graphicx}
\newtheorem{theorem}{Theorem}
\usepackage{url}

\usepackage{tabularx,booktabs}
\newcolumntype{C}{>{\centering\arraybackslash}X} 
\usepackage{lipsum}

\usepackage{makecell} 

\usepackage{graphicx}
\usepackage{color}

\newtheorem{lem}{Lemma}

\newtheorem{corollary}{Corollary}

\definecolor{dblue}{RGB}{15,89,164}

\begin{document}
\title { Active RIS  Versus Passive RIS: Which Is Superior with the Same Power Budget?}
\author{Kangda Zhi, Cunhua Pan, Hong Ren, Kok Keong Chai, and Maged Elkashlan\thanks{\itshape (Corresponding author: Cunhua Pan.)\upshape
		 		
		K. Zhi, K. K. Chai, and M. Elkashlan are with the School of Electronic Engineering and Computer Science at Queen Mary University of London, UK. (e-mail: k.zhi, michael.chai, maged.elkashlan@qmul.ac.uk).
		
		C. Pan, H. Ren are with the National Mobile Communications Research Laboratory, Southeast University, China. (cunhuapan21@gmail.com, hren@seu.edu.cn).}
}

\maketitle

\begin{abstract}
This letter theoretically compares the active reconfigurable intelligent surface (RIS)-aided system with the passive RIS-aided system. 
For fair comparison, we consider that these two systems have the same overall power budget that can be used at both the base station (BS) and the RIS. For active RIS, we first derive the optimal power allocation between the BS's transmit signal power and RIS's output signal power. We also analyze the impact of various system parameters on the optimal power allocation ratio. Then, we compare the performance between the active RIS and the passive RIS, which demonstrates that the active RIS would be superior if the power budget is not very small and the number of RIS elements is not very large.
\end{abstract}


\IEEEpeerreviewmaketitle

\section{Introduction}
Passive reconfigurable intelligent surface (RIS)-aided systems have attracted extensive research attention recently\cite{di2020smart,wu2019intelligent,pan2020intelligent,huang2019reconfigu,pan2020multicell}. By passively reflecting the impinging signal and intelligently adjusting the phase shifts, received signals from different paths can be constructively superimposed and enhanced. Meanwhile, due to its passive nature, nearly zero additional power is needed, which is promising and attractive for next-generation communication systems. 

However, the passive nature also has some drawbacks. The signal reflected by the RIS needs to pass through two paths, i.e., the base station (BS)-RIS and RIS-user paths. Without signal amplification, the received signal suffers from the product/double path-loss attenuation and therefore becomes weak enough. This ``double path-loss'' attenuation limits the potential of RIS to a large extent\cite{bigison2020IRS}. To tackle this challenge, the active RIS equipped with the power amplification capability has been proposed and investigated in\cite{zhang2021active,long2021active,you2021active,Majid2021active,liu2021active,zeng2021active,xu2021resource}. Integrated with reflection-type amplifiers\cite{zhang2021active}, the active RIS can not only adjust the phase shifts but also amplify the received signal attenuated from the first hop to a normal strength level. As a result, active RIS effectively circumvents the double path-loss attenuation and fully unleashes the potential of the RIS. Besides, it is worth noting that the active RIS does not have radio-frequency chains, and therefore it is fundamentally different from the conventional relay\cite{zhang2021active,long2021active}.

Compared with the passive RIS, a noticeable feature of active RIS is that it needs additional power to supply the amplifiers. Therefore, the active RIS requires a larger power budget than the passive RIS given the same number of reflecting elements. Even though the superiority of the active RIS over the passive RIS has been widely confirmed in \cite{zhang2021active,long2021active,you2021active,Majid2021active,liu2021active,zeng2021active,xu2021resource}, none of them has performed a theoretical analysis considering the constraint that the same overall power budget is employed for active and passive RISs, which motivates our study.

In this letter, we perform a fair comparison between the active and passive RISs with the same overall power budget. Given total power, we first decide how much power should be allocated to the active RIS's amplifiers. Then, we theoretically analyze the impact of various system parameters on the derived optimal power allocation ratio. Finally, analytical and numerical results are provided to shed light on the performance difference between the active and passive RISs.

\section{System Model}
\begin{figure}
	\setlength{\abovecaptionskip}{0pt}
	\setlength{\belowcaptionskip}{-20pt}
	\centering
	\includegraphics[width= 0.3\textwidth]{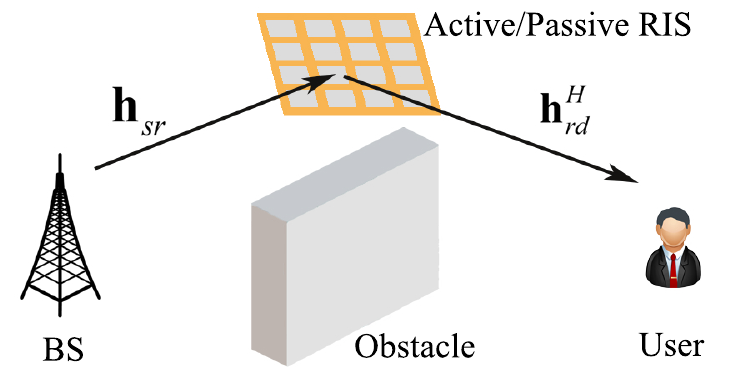}
	\DeclareGraphicsExtensions.
	\caption{Illustration of the considered system.}
	\label{figure1}
	\vspace{-10pt}
\end{figure}

We consider a single-input single-output (SISO) system with the aid of an RIS equipped with $ N $ elements as illustrated in Fig. \ref{figure1}. The BS-RIS and RIS-user channels are denoted by $\mathbf{h}_{sr}\in \mathbb{C}^{N\times 1}$ and $\mathbf{h}_{rd}^H\in \mathbb{C}^{1\times N}$, respectively. For the sake of comparison, we assume that the direct channel is blocked due to obstacles.

As in \cite{you2021active,long2021active}, we assume that line-of-sight (LoS) paths exist in RIS-reflected channels. Based on the uniform linear array (ULA) model, we denote that $\mathbf{h}_{s r}=h_{s r}[1, e^{-j \frac{2 \pi}{\lambda} \vartheta}, \ldots, e^{-j \frac{2 \pi}{\lambda}(N-1) \vartheta}]^{T}$ and $\mathbf{h}_{r d}^{H}=h_{r d}[1, e^{-j \frac{2 \pi}{\lambda} \varsigma}, \ldots, e^{-j \frac{2 \pi}{\lambda}(N-1) \varsigma}]$,
where $\vartheta$ and $\varsigma$ are the angles of arrival and departure, respectively. $h_{sr}$ and $h_{rd}$ represent the distance-dependent path-loss factors expressed as $ h_{s r}=\sqrt{\beta_{0} d_{s r}^{-\alpha_{s r}}} $ and $ h_{r d}=\sqrt{\beta_{0} d_{r d}^{-\alpha_{r d}}} $, where $\beta_{0}$ represents the reference strength for the channel at a distance of $1$ m, $d_{sr}$ and $d_{rd}$ are distances, and $\alpha_{s r}$ and $\alpha_{r d}$ denote the path-loss exponents. Note that it is general to set $\beta_{0}=-30$ dB and $2\leq \alpha_{s r },\alpha_{r d}\leq 4$. Accordingly, for moderate distances of $ d_{s r },d_{r d}= 100$ m and path exponents of $\alpha_{s r },\alpha_{r d}= 2$, the channel gains would be $h_{s r }^2,h_{r d}^2= -70$ dB, which are very small values. Without amplification, the received signal at the user suffers from the product/double path-loss $h_{s r }^2\times h_{r d}^2= -140$ dB and therefore becomes very weak. In the sequel of this letter, the order of magnitude of $h_{s r }^2$ and $h_{rd }^2$ could help us better understand the performance comparison between the active RIS and the passive RIS.

Define the reflection matrix of the RIS as $\mathbf{\Theta}=\operatorname{diag}\left\{\rho_{1} e^{j \theta_{1}}, \ldots, \rho_{N} e^{j \theta_{N}}\right\}$, where $\theta_n$ denotes the phase shift of the $n$-th RIS reflecting element. For passive RIS, we have $\rho_n=1,\forall n$. However, $\rho_n>1,\forall n$ are feasible for active RIS due to its amplifiers. For simplicity, we assume that $\rho_n=\rho,\forall n$ and then define $\boldsymbol{\Theta}=\rho \operatorname{diag}\left\{e^{j \theta_{1}}, \ldots, e^{j \theta_{N}}\right\}=\rho \boldsymbol{\Phi}$. Let $x \sim \mathcal{C} N\left(0, 1\right)$ denote the symbol transmitted from the BS. Then, the signal reflected by the active RIS and finally received by the user is given by
\begin{align}\label{signal_active}
y_\mathrm{a c t}=\sqrt{P_\mathrm{B S}^\mathrm{a c t}} \rho \mathbf{h}_{r d}^{H} \mathbf{\Phi} \mathbf{h}_{s r} x+\rho \mathbf{h}_{r d}^{H} \mathbf{\Phi} \mathbf{n}_{r}+n,
\end{align}
where $ P_\mathrm{B S}^\mathrm{a c t} $ is the transmit power of BS in active RIS systems, $\mathbf{n}_{r} \sim \mathcal{C N}\left(\mathbf{0}, \sigma_{r}^{2} \mathbf{I}_{N}\right)$ denotes the thermal noise introduced by active RIS components, and $n \sim \mathcal{C} N\left(0, \sigma^{2}\right)$ represents the thermal noise at the receiver. Letting $\rho=1$ and $\mathbf{n}_r=\mathbf{0}$, we can obtain the received signal in the passive RIS-aided system as follows
\begin{align}
	y_\mathrm{pas}=\sqrt{P_\mathrm{B S}^\mathrm{pas}}  \mathbf{h}_{r d}^{H} \mathbf{\Phi} \mathbf{h}_{s r} x+n,
\end{align}
where $ P_\mathrm{B S}^\mathrm{pas} $ is the transmit power of BS in passive RIS systems. Besides, the overall power consumption of  active and passive RIS-aided systems are respectively given by\cite{long2021active}
\begin{align}\label{power_model}
\begin{aligned}
	&Q_{\text {act }}=P_{\mathrm{BS}}^{\text {act }}+P_{\mathrm{RIS}}^{\mathrm{act}}+N\left(P_{\mathrm{SW}}+P_{\mathrm{DC}}\right), \\
	&Q_{\mathrm{pas}}=P_{\mathrm{BS}}^{\mathrm{pas}}+N P_{\mathrm{SW}},
\end{aligned}
\end{align}
where $ P_{\mathrm{RIS}}^{\mathrm{act}} $ is the output signal power of active RIS, $ P_{\mathrm{SW}} $ is the power consumed by the phase shift switch and control circuit in each RIS elements, $ P_{\mathrm{DC} }$ is the direct current biasing power used by each active RIS element.

\section{Active RIS versus Passive RIS}
Based on (\ref{signal_active}), the achievable rate of active RIS systems is $R_\text{act} = \log_2 \left(1+ \gamma_{\text{act}}\right)$, in which the signal-to-noise-ratio (SNR) is expressed as
\begin{align}\label{SNR_active}
	\gamma_{\text {act }}&=\frac{P_{\mathrm{BS}}^{\text {act }} \rho^{2}\left|\mathbf{h}_{r d}^{H} \boldsymbol{\Phi} \mathbf{h}_{s r}\right|^{2}}{\rho^{2} \sigma_{r}^{2}\left\|\mathbf{h}_{r d}^{H} \boldsymbol{\Phi}\right\|^{2}+\sigma^{2}}
	{=}\frac{P_{\mathrm{BS}}^{\text {act }} \rho^{2}\left|\mathbf{h}_{r d}^{H} \mathbf{\Phi} \mathbf{h}_{s r}\right|^{2}}{\rho^{2} \sigma_{r}^{2} N h_{r d}^{2}+\sigma^{2}}\nonumber\\
	&\overset{(a)}{=}\frac{P_{\mathrm{BS}}^{\mathrm{act}} \rho^{2} N^{2} h_{s r}^{2} h_{r d}^{2}}{\rho^{2} \sigma_{r}^{2} N h_{r d}^{2}+\sigma^{2}}
	=\frac{N^{2} P_{\mathrm{BS}}^{\mathrm{act}} h_{s r}^{2} h_{r d}^{2}}{N \sigma_{r}^{2} h_{r d}^{2}+\frac{\sigma^{2}}{\rho^{2}}},
\end{align}
where $(a)$ utilizes the optimal design of $\mathbf{\Phi}$, i.e., $\theta_{n}=\arg \left\{\left[\mathbf{h}_{rd}\right]_{n}\right\}-\arg \left\{\left[\mathbf{h}_{sr}\right]_{n}\right\}, \forall n  $.

Substituting (\ref{SNR_active}) with $\rho=1$ and $\sigma_{r}^2=0$, we obtain the achievable rate of passive RIS systems as $R_\mathrm{pas} = \log_2 \left(1+ \gamma_{\mathrm{pas}}\right)$, where
\begin{align}\label{SNR_passive}
	\gamma_{\text {pas }}=\frac{N^{2} P_{\mathrm{BS}}^{\mathrm{pas}} h_{s r}^{2} h_{r d}^{2}}{{\sigma^{2}}}.
\end{align}

Comparing (\ref{SNR_active}) with (\ref{SNR_passive}), it is obvious that $\gamma_{\text {act }}> \gamma_{\text {pas }}$, if $P_{\mathrm{BS}}^{\text {act }}=P_{\mathrm{BS}}^{\text {pas }}$ (without the same power budget constraint), $N h_{r d}^{2}\ll 1$ (for $h_{r d}^{2}\approx  -70$ dB), $\sigma_{r}^{2} \approx \sigma^{2}$, and $\rho^2> 1$. Therefore, for a fair comparison, it is necessary to constrain that both two schemes have the same overall power budget $Q_{\text {act }}=Q_{\text {pas }}$, which means $P_{\mathrm{BS}}^{\text {act }}<P_{\mathrm{BS}}^{\text {pas }}$. In this context, the superiority of active RIS over passive RIS is non-trivial and needs to be re-examined again.

\subsection{Problem Formulation with the Same Power Budget}
Assume that the total power budget is $Q_\mathrm{tot}$, i.e., $ Q_{\text {act }}=Q_{\text {pas }} =Q_\mathrm{tot}$. From (\ref{power_model}), we have \begin{align}
		&P_{\mathrm{BS}}^{\mathrm{act}}=Q_{\mathrm{tot}}-N\left(P_{\mathrm{SW}}+P_{\mathrm{DC}}\right)-P_{\mathrm{RIS}}^{\text {act }} \triangleq C-P_{\mathrm{RIS}}^{\text {act }}, \\
		&P_{\mathrm{BS}}^{\mathrm{pas}}=Q_{\mathrm{tot}}-N P_{\mathrm{SW}}=C+N P_{\mathrm{DC}},
\end{align}
where $C= P_{\mathrm{BS}}^{\mathrm{act}}+P_{\mathrm{RIS}}^{\mathrm{act}}$ corresponds to the available power left for allocating to the BS and active RIS after supplying the hardware power consumption. Clearly, if $C\leq0$, we have $P_{\mathrm{BS}}^{\mathrm{act}}=0$ and $\gamma_{\text {act }}=0$. Therefore, we only focus on the region of $C> 0$ in this section.

Different from the passive RIS, we need to additionally decide the optimal power allocation for $P_{\mathrm{BS}}^{\mathrm{act}}$ and $P_{\mathrm{RIS}}^{\mathrm{act}}$ given $C$. The optimization problem is formulated as
\begin{subequations}
	\begin{align}\label{Problem1}
&\max _{P_{\mathrm{BS}}^{\text {act }}, \; \rho}  \;\;\;  \;  \;  \gamma_{\text {act }} \\\label{10b}
&\text { s.t. }  \qquad P_{\mathrm{BS}}^{\text {act }} \rho^{2}   \|{\bf\Phi} \mathbf{h}_{sr}\|^{2}+\rho^{2} \sigma_{r}^{2}\|\boldsymbol{\Phi}\|^{2}\leq P_{\mathrm{RIS}}^{\text {act }}, \\\label{10c}
&\qquad\qquad P_{\mathrm{BS}}^{\text {act }} + P_{\mathrm{RIS}}^{\text {act }}=C.
	\end{align}
\end{subequations}


Since $ \gamma_{\text {act }}$ increases with $\rho$, substituting (\ref{10c}) into (\ref{10b}), the optimal $\rho^2$ should satisfy the following condition
\begin{align}\label{rho_condition}
	\rho^{2} = \frac{C-P_{\mathrm{BS}}^{\mathrm{act}}}{N\left(P_{\mathrm{BS}}^{\mathrm{act}} h_{sr}^{2}+\sigma_{r}^{2}\right)}.
\end{align}
Then, the original problem is transformed to
\begin{subequations}
	\begin{align}\label{snr_active_objective}
		&\max _{P_{\mathrm{BS}}^{\mathrm{act }}}  \;  \;  \;  \gamma_\mathrm{a c t}=\frac{N h_{s r}^{2} h_{r d}^{2}\left(C P_{\mathrm{BS}}^{\text {act }}-(P_{\mathrm{BS}}^{\text {act } }) ^2    \right)}{\sigma_{r}^{2} h_{r d}^{2}\left(C-P_{\mathrm{BS}}^{\text {act }}\right)+\sigma^{2}\left(P_{\mathrm{BS}}^{\text {act }} h_{sr}^{2}+\sigma_{r}^{2}\right)} \\\label{11b}
		&\text { s.t. }  \qquad 0\leq  P_{\mathrm{BS}}^{\text {act }} \leq C.
	\end{align}
\end{subequations}

It is readily to find $\left.     \gamma_\mathrm{a c t} \right|_ {P_{\mathrm{BS}}^{\mathrm{act }} = 0} = 0$ and  $\left.     \gamma_\mathrm{a c t} \right|_ {P_{\mathrm{BS}}^{\mathrm{act }} =C} =0$. Therefore, it is necessary to decide the optimal power allocation between the BS and the active RIS. 

\subsection{Optimal Power Allocation}
\begin{theorem}
	If $\sigma_{r}^{2} h_{rd}^{2}=\sigma^{2} h_{sr}^{2}$, the optimal power allocated to the BS is $(P_\mathrm{B S}^\mathrm{a c t})^{\star}={C}/{2}$. Otherwise, we have
	\begin{align}\label{optimal_power_allocation}
	\begin{aligned}
		\left(P_{\mathrm{BS}}^{\mathrm{act}}\right)^{\star}&=\frac{1}{\sigma_{r}^{2} h_{r d}^{2}-\sigma^{2} h_{s r}^{2}} \times \Big\{C \sigma_{r}^{2} h_{r d}^{2}+\sigma^{2} \sigma_{r}^{2} \\
		&-\sqrt{\left(C \sigma^{2} h_{s r}^{2}+\sigma^{2} \sigma_{r}^{2}\right)\left(C \sigma_{r}^{2} h_{r d}^{2}+\sigma^{2} \sigma_{r}^{2}\right)}\;\Big\}.
	\end{aligned}
	\end{align}

Meanwhile, the optimal power allocated to the RIS is $\left(P_\mathrm{RIS}^\mathrm{a c t}\right)^{\star}=C-\left(P_\mathrm{B S}^\mathrm{a c t}\right)^{\star}$.
\end{theorem}

\itshape {Proof:}  \upshape Please refer to Appendix \ref{App_1}. \hfill $\blacksquare$

If $\sigma_r^{2}= \sigma^{2}$, condition $\sigma_{r}^{2} h_{rd}^{2}=\sigma^{2} h_{sr}^{2}$ can be satisfied when the RIS is located in the middle between the BS and the user. In this special case, it is optimal to equally allocate the power to the BS and the RIS. Unless otherwise stated, we focus on the case $ \sigma_{r}^{2} h_{r d}^{2} \neq \sigma^{2} h_{s r}^{2}$ in the following.

\begin{corollary}\label{corollary3}
	When $\sigma_{r}^2 \to 0$, $\left(P_\mathrm{B S}^\mathrm{a c t}\right)^{\star}  \to 0$ and $\left(P_\mathrm{RIS}^\mathrm{a c t}\right)^{\star} \to C$. When $\sigma_{r}^2 \to \infty$, $\left(P_\mathrm{B S}^\mathrm{a c t}\right)^{\star} \to C+\frac{\sigma^{2}}{h_{r d}^{2}}-\sqrt{\frac{\sigma^{2}}{h_{r d}^{2}}\left(C+\frac{\sigma^{2}}{h_{r d}^{2}}\right)}$, which further tends to $C$ if $\frac{\sigma^{2}}{h_{r d}^{2}} \ll C  $.
\end{corollary}

As $\sigma_{r}^2 \to 0$, the receiver noise $\sigma^2$ becomes dominant, and therefore larger $P_\mathrm{RIS}^\mathrm{a c t}$ is preferred, which achieves larger $\rho$ and then reduces the term $\frac{\sigma^2}{\rho^2}$ in (\ref{SNR_active}). As $\sigma_{r}^2 \to \infty$, the RIS noise becomes dominant, and it is useless to increase $\rho$ since it also amplifies the RIS noise term $\rho^2\sigma_{r}^2N h_{rd}^2$ in (\ref{SNR_active}). In this case, increasing $P_\mathrm{B S}^\mathrm{a c t}$ can effectively improve the SNR.

\begin{corollary}\label{corollary1}
	Both $ \left(P_\mathrm{B S}^\mathrm{a c t}\right)^{\star}  $ and $\left(P_\mathrm{RIS}^\mathrm{a c t}\right)^{\star}$ are increasing functions of $Q_\mathrm{tot}$ but decreasing functions of $N$. 
\end{corollary}

\itshape {Proof:}  \upshape Please refer to Appendix \ref{App_2}. \hfill $\blacksquare$

Corollary \ref{corollary1} shows that as power budget $Q_\mathrm{tot}$ grows, it is optimal to simultaneously increase the power of the BS and the RIS.  Fully allocating the increased power to the BS/RIS unfairly will sacrifice the performance. Meanwhile, when $N$ increases, less power is left for the BS and the RIS, and it is optimal to simultaneously cut down their power. 

\begin{corollary}\label{corollary2}
	When $\sigma_{r}^{2} h_{rd}^{2} > \sigma^{2} h_{sr}^{2}$, $ \left(P_\mathrm{B S}^\mathrm{a c t}\right)^{\star}  - \left(P_\mathrm{RIS}^\mathrm{a c t}\right)^{\star}$ is an increasing function of $Q_\mathrm{tot}$ but decreasing function of $N$. On the contrary, when $\sigma_{r}^{2} h_{rd}^{2} < \sigma^{2} h_{sr}^{2}$, $ \left(P_\mathrm{B S}^\mathrm{a c t}\right)^{\star}  - \left(P_\mathrm{RIS}^\mathrm{a c t}\right)^{\star}$ is a decreasing function of $Q_\mathrm{tot}$ but increasing function of $N$. 
\end{corollary}

\itshape {Proof:}  \upshape Please refer to Appendix \ref{App_3}. \hfill $\blacksquare$

Although the power should be allocated fairly, Corollary \ref{corollary2} unveils that there do exist some allocation priorities for increased $Q_\mathrm{tot}$. With larger $\sigma_{r}^{2} h_{rd}^{2}$, more power should be allocated  to the BS, since $ P_\mathrm{B S}^\mathrm{a c t} $ reduces the impact of RIS noise. By contrast, with larger $ \sigma^{2} h_{sr}^{2}$, more power should be allocated to the active RIS, which leads to larger $\rho$ and therefore decreases the impact of receiver noise. Besides, by moving the RIS closer to the user, $h_{rd}^2$ increases while $ h_{sr}^2 $ decreases. Hence, when the RIS is located near the BS (user), more power should be allocated to the RIS (BS). 
This is because the received signal at the RIS is stronger with larger $h_{sr}^2$, and larger $ P_\mathrm{RI S}^\mathrm{a c t}$ is needed to amplify a stronger signal for a certain multiple $\rho$ as shown in (\ref{rho_condition}).

\subsection{SNR Comparison}
Using $\left(P_{\mathrm{BS}}^{\mathrm{act}}\right)^{\star}$, (\ref{SNR_active}), (\ref{SNR_passive}), and (\ref{rho_condition}), by solving $\gamma_{\text {act }}>\gamma_{\text {pas}}$, it is readily to obtain the following results.
\begin{lem}
	Passive RIS performs better if
	\begin{align}\label{condition1}
 N h_{r d}^{2} \frac{\sigma_{r}^{2}}{\sigma^{2}} >\frac{(P_{\mathrm{BS}}^{\mathrm{act}})^{\star}}{P_{\mathrm{BS}}^{\mathrm{pas}}}.
	\end{align}
	 Otherwise, active RIS performs better when 
	\begin{align}\label{condition2}
 \frac{C-(P_{\mathrm{BS}}^{\mathrm{act}})^ {\star}      }{N\left(     \left(P_{\mathrm{BS}}^{\mathrm{act}}\right)^{\star} h_{sr}^{2}+\sigma_{r}^{2}\right)}=\left(\rho^{\star}\right)^{2}>\frac{1}{\frac{(P_{\mathrm{BS}}^{\mathrm{act}})^{\star}}{P_{\mathrm{BS}}^{\mathrm{pas}}}-N h_{r d}^{2} \frac{\sigma_{r}^{2}}{\sigma^{2}}}.
	\end{align}
\end{lem}

We firstly focus on condition (\ref{condition1}).
Since $ \frac{(P_{\mathrm{BS}}^{\mathrm{act}})^{\star}}{P_{\mathrm{BS}}^{\mathrm{pas}}}<1$, passive RIS must be better if $N>\frac{\sigma^{2}}{h_{r d}^{2} \sigma_{r}^{2}}$, i.e., for sufficiently large $N$, since in this case active RIS suffers from severe noise $N \sigma_{r}^{2} h_{r d}^{2}$. (\ref{condition1}) is easier to hold for small $\sigma^2$ or large $\sigma_{r}^2$. This is because active RIS can effectively mitigate the impact of $\sigma^2$ but it is impaired by $\sigma_{r}^2$. If $\sigma_{r}^2=\sigma^2$, condition (\ref{condition1}) is equal to $N h_{r d}^{2}>1$, which means that the attenuation from path-loss $h_{r d}^{2}$ is compensated by RIS's gain $N$ and therefore double path-loss attenuation no longer exists. However, we emphasize that the $N$ satisfying condition (\ref{condition1}) may be very large due to the small value of $h_{r d}^{2}$.

\begin{lem}\label{lem_P_ratio}
Define $g(C)=\frac{(P_{\mathrm{BS}}^{\mathrm{act}})^{\star}}{P_{\mathrm{BS}}^{\mathrm{pas}}}$.  
If $\sigma_{r}^{2}  h_{r d}^{2} > \sigma_{r}  h_{sr }^{2}$,  $ g(C)$ increases with $C$.
If $\sigma_{r}^{2}  h_{r d}^{2} < \sigma_{r}  h_{sr }^{2}$,  depending on the values of $N$, $ g(C)$ could be an increasing function or a function which firstly increases but then decreases with $C$.
 In addition, $g(0)\to0$ and $ g(\infty)  \to    \frac{\sigma_{r}^{2} h_{r d}^{2}\! -\! \sqrt{\sigma^{2} h_{s r}^{2} \sigma_{r}^{2} h_{r d}^{2}}}{\sigma_{r}^{2} h_{r d}^{2}-\sigma^{2} h_{s r}^{2}}\in (0,1)$ which approaches $1$ as $\sigma_{r}^{2}  h_{r d}^{2} \rightarrow \infty$.
\end{lem}

\itshape {Proof:}  \upshape Please refer to Appendix \ref{App_4}. \hfill $\blacksquare$

\begin{corollary}\label{corollary4}
	Passive RIS outperforms active RIS for small $C$, i.e., for low power budget $Q_\mathrm{tot}$.
\end{corollary}

\itshape {Proof:}  \upshape  
Define $C^{*}=\underset{C}{\arg \max } \;g(C)$. 
If $g(C^*)<N h_{r d}^{2} \frac{\sigma_{r}^{2}}{\sigma^{2}} $, condition (\ref{condition1}) holds for all $C$ (also for small $C$).
If not, from Lemma \ref{lem_P_ratio}, there exists an intersection point $\widetilde{C}\in (0,C^*)$ so that $g(\widetilde{C})=N h_{r d}^{2} \frac{\sigma_{r}^{2}}{\sigma^{2}} $.
Then, condition (\ref{condition1}) holds for all $C\in (0,\widetilde{C})$.
 \hfill $\blacksquare$

The above discussions demonstrate that passive RIS is better for very large $N$ and very small $Q_\mathrm{tot}$. We next use condition (\ref{condition2}) to demonstrate the superiority of active RIS when $N$ is not very large and $Q_\mathrm{tot}$ is not very small.
\begin{corollary}\label{corollary5}
	Active RIS outperforms passive RIS if $\sigma^2\approx \sigma_{r}^2\ll P_\mathrm{DC}$, $Nh_{sr}^2, Nh_{rd}^2 \ll 1$, and $C\geq NP_\mathrm{DC}$.
\end{corollary}

\itshape {Proof:}  \upshape We prove $\gamma_{\text {act }}((P_{\mathrm{BS}}^{\mathrm{act}})^{\star})>  \gamma_{\text {pas }}$ by proving $\gamma_{\text {act }}(C/2)>  \gamma_{\text {pas }}$ due to $\gamma_{\text {act }}((P_{\mathrm{BS}}^{\mathrm{act}})^{\star}) \geq  \gamma_{\text {act }}(C/2)$. By replacing $(P_{\mathrm{BS}}^{\mathrm{act}})^{\star}$ in (\ref{condition2}) with $C/2$, the left-hand side of (\ref{condition2}) is lower bounded by
\begin{align}\label{condition_left_approx}
\rho_\mathrm{sub}^{2}& = \frac{C}{C N h_{s r}^{2}+2 N \sigma_{r}^{2}}\geq \min \left\{\frac{1}{2 N h_{s r}^{2}}, \frac{C}{4 N \sigma_{r}^{2}}\right\} \nonumber\\
&\geq \min \left\{\frac{1}{2 N h_{s r}^{2}}, \frac{P_\mathrm{DC}}{4 \sigma_{r}^{2}}\right\} \gg 1.
\end{align}
The right-hand side of (\ref{condition2}) is approximately upper bounded by
\begin{align}\label{condition_right_appro}
\frac{1}{\frac{C}{2\left(C  +  N  P_{\mathrm{DC}}\right)}- N h_{r d}^{2} \frac{\sigma_{r}^{2}}{\sigma^{2}}}  \leq    \frac{1}{\frac{1}{4}  -     N h_{r d}^{2} \frac{\sigma_{r}^{2}}{\sigma^{2}}}  \approx  4<\rho_\mathrm{sub}^2,
\end{align}
which completes the proof.
\hfill $\blacksquare$

The reason behind Corollary \ref{corollary5} is that the signal received at the RIS has been attenuated by path-loss $h_{sr}^2$ and therefore becomes very weak. According, large $\rho^2$ is feasible to amplify this weak signal to a normal strength while satisfying the RIS power constraint (\ref{10b}).

\section{Simulation Results}
We consider the same simulation setup as in \cite{long2021active,zhang2021active}. If not specified otherwise, we set $N=256$, $P_{\mathrm{DC}}=-5$ dBm, $P_{\mathrm{SW}}=-10$ dBm, $\sigma^2=\sigma_{r}^2=-70$ dBm, and $Q_\mathrm{tot}=30$ dBm. For the path-loss factors, we set $\beta_{0}=-30$ dB,  $\alpha_{s r}=\alpha_{r d}=2$.
The BS and the user are located at $(0,0,0)$ and $(100\;\mathrm{m},0,0)$, respectively. The RIS is located at $ (x_\mathrm{RIS},0,10 \;\mathrm{m})$ where $x_\mathrm{RIS}=90$ m is adopted by default.

We first validate our conclusion in Corollary \ref{corollary5}. Based on the above simulation setup and use the sub-optimal solution $P_{\mathrm{BS}}^{\mathrm{act}}=C/2$, we can calculate the left- and right-hand side of (\ref{condition2}) as $\rho_\mathrm{sub}^{2}= 3.2\times10^4$ and $2.18$, respectively, which verifies the correctness of Corollary \ref{corollary5}.

\begin{figure}[t]
	\setlength{\abovecaptionskip}{0pt}
	\setlength{\belowcaptionskip}{-20pt}
	\centering
	\includegraphics[width= 0.4\textwidth]{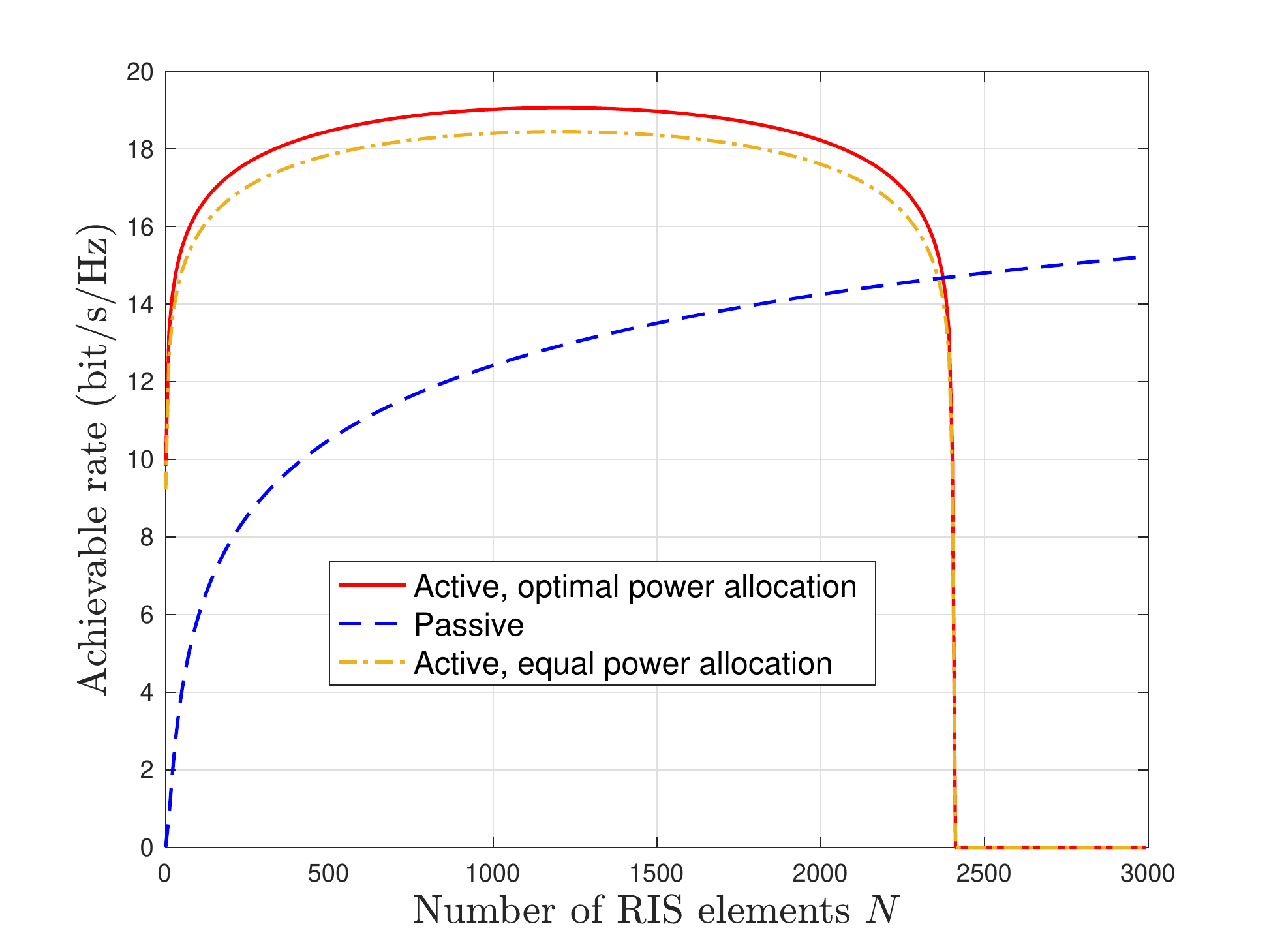}
	\DeclareGraphicsExtensions.
	\caption{Rate comparison versus $N$.}
	\label{figure2}
	\vspace{-10pt}
\end{figure}

Fig. \ref{figure2} illustrates the superiority of active RIS when $N$ is not very large. This is because active RIS can exploit a small amount of power to amplify the signal attenuated after the transmission in the first hop, and therefore significantly improve the strength of the signal received at the user. When $N$ is very large, passive RIS becomes superior while the data rate of active RIS begins to decrease. This is because a huge amount of power is consumed by active RIS to supply its amplifiers and also because of the stronger active RIS thermal noise $Nh_{rd}^2\sigma_{r}^2$. To sum up, active RIS is more promising than passive RIS since it can achieve a high achievable rate with a small number of elements.

\begin{figure}[t]
	\setlength{\abovecaptionskip}{0pt}
	\setlength{\belowcaptionskip}{-20pt}
	\centering
	\includegraphics[width= 0.4\textwidth]{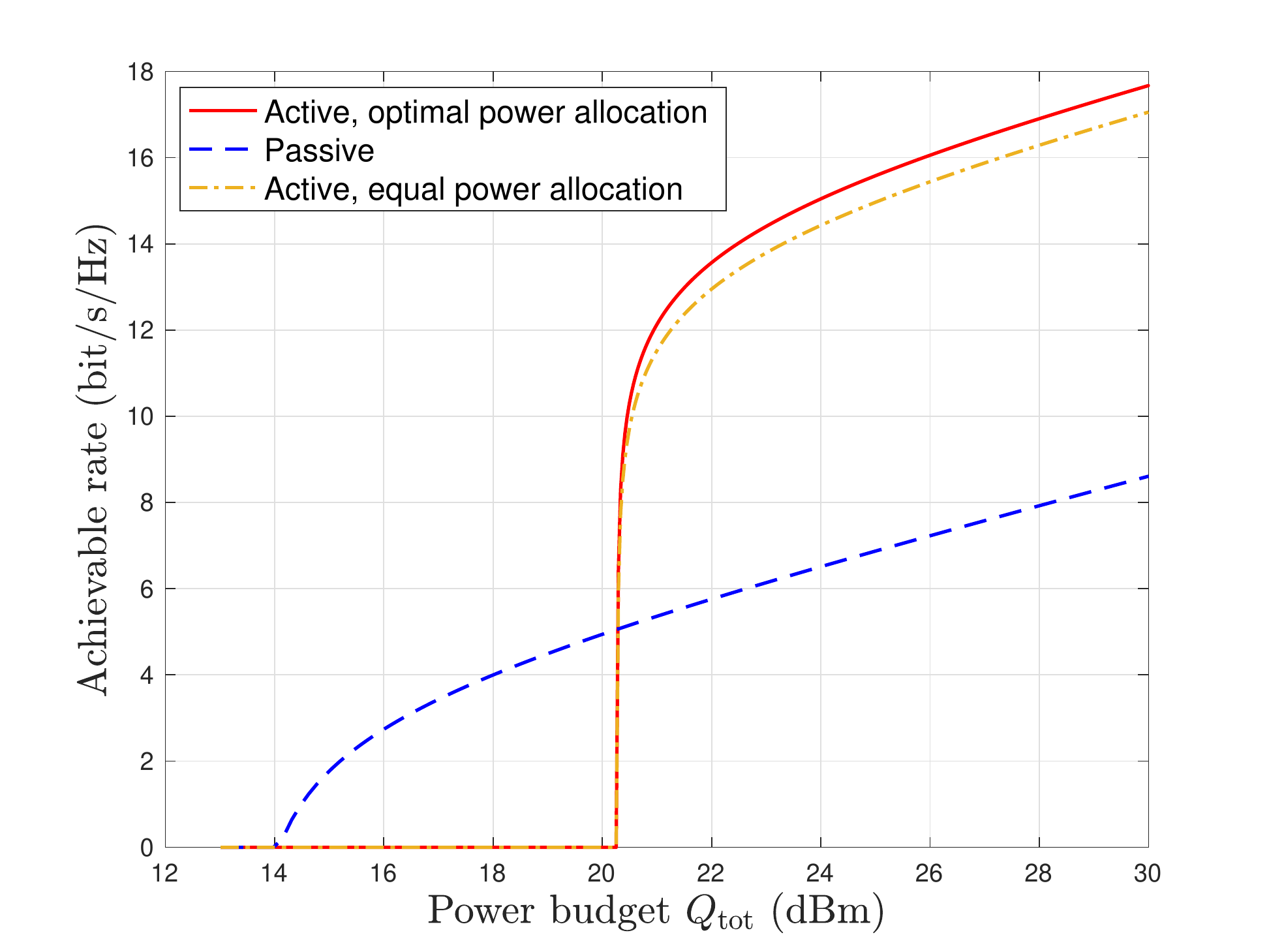}
	\DeclareGraphicsExtensions.
	\caption{Rate comparison versus power budget $Q_\mathrm{tot}$.}
	\label{figure3}
	\vspace{-10pt}
\end{figure}

Being consistent with Corollaries \ref{corollary4} and \ref{corollary5}, Fig. \ref{figure3} unveils that the passive RIS performs better for small power budgets while the active RIS is superior when the power budget is sufficient. Meanwhile, it can be seen that the equal allocation strategy begins to show its defects as $Q_\mathrm{tot}$ increases. This is because as the power budget grows, the optimal allocation ratio between the BS and active RIS needs to be adjusted accordingly, as discussed in Corollary \ref{corollary2}.

\begin{figure}[t]
	\setlength{\abovecaptionskip}{0pt}
	\setlength{\belowcaptionskip}{-20pt}
	\centering
	\includegraphics[width= 0.4\textwidth]{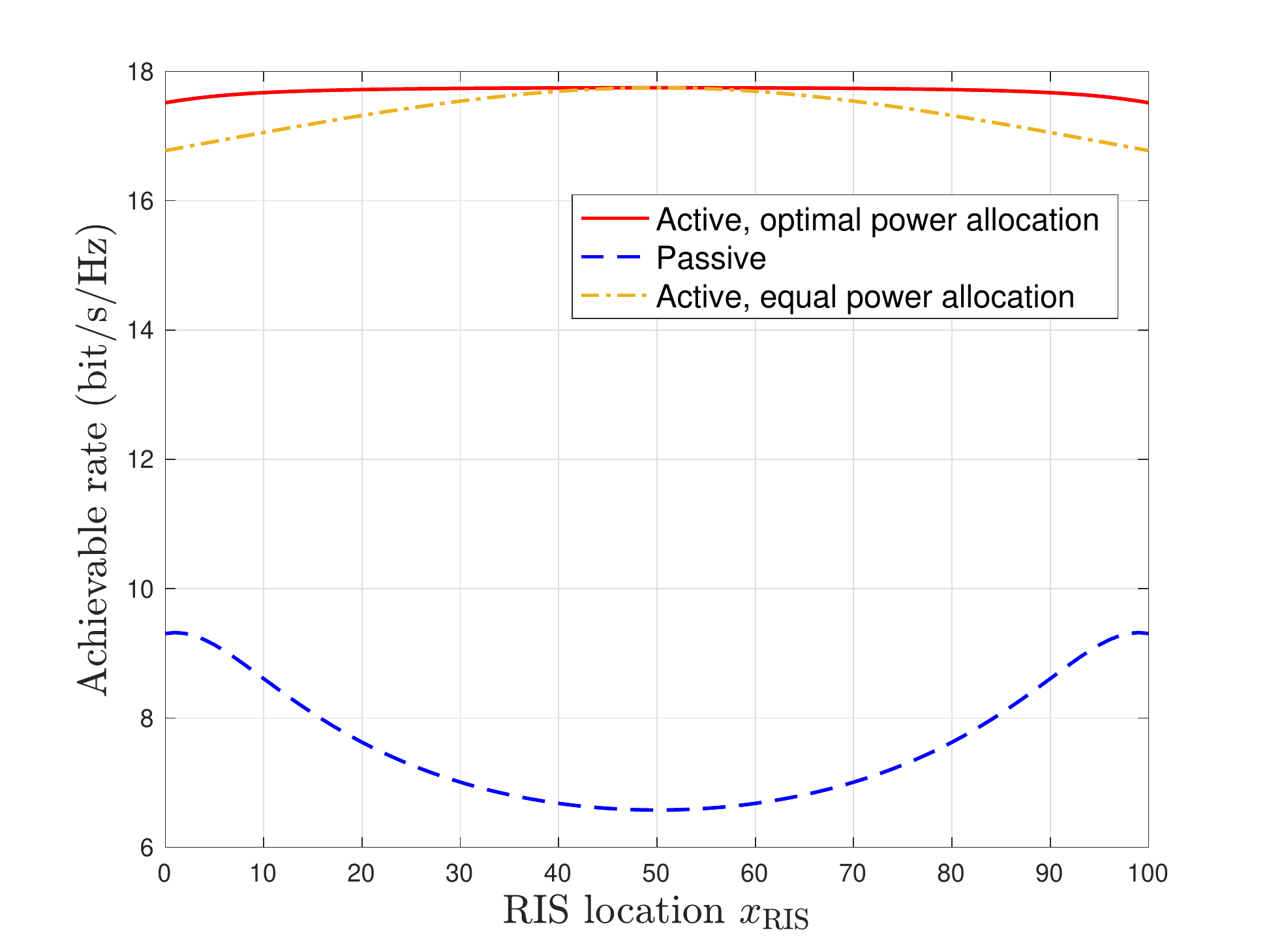}
	\DeclareGraphicsExtensions.
	\caption{Rate comparison versus RIS location $x_\mathrm{RIS}$.}
	\label{figure4}
	\vspace{-10pt}
\end{figure}

Finally, Fig. \ref{figure4} compares the active and passive RISs for different deployments. It is shown that active RIS can perform constantly better in all deployments. It is known that the passive RIS should be deployed near the BS or the user in order to alleviate the attenuation from double path-loss. However, the active RIS can be deployed flexibly since the double-fading effect is effectively circumvented thanks to the integration of amplifiers. Besides, compared to the equal power allocation scheme which only achieves optimality for a middle deployment, the proposed optimal allocation scheme can dynamically balance the power allocation and then make the achievable rate of active RIS stable in all locations.

\section{Conclusion}
This letter theoretically compared the active RIS with the passive RIS under the same power budget. We firstly derived the optimal power allocation ratio between the BS and the active RIS. We then provided some insights based on the derived allocation ratio and discussed the conditions when active or passive RIS performs better.


\begin{appendices}
\section{}\label{App_1}
The first-order derivative of $\gamma_{\text {act }}$ in (\ref{snr_active_objective}) with respect to $P_\mathrm{B S}^\mathrm{a c t}$ is given by
\begin{align}\label{snr_derivative}
\frac{\partial \gamma_{\text {act }}}{\partial P_{\mathrm{BS}}^{\text {act }}}=\frac{  N h_{s r}^{2} h_{r d}^{2}  \times f_{1}\left(P_{\mathrm{BS}}^{\text {act }}\right)}{\left(\sigma_{r}^{2} h_{r d}^{2}\left(C-P_{\mathrm{BS}}^{\text {act }}\right)+\sigma^{2}\left(P_{\mathrm{BS}}^{\text {act }} h_{s r}^{2}+\sigma_{r}^{2}\right)\right)^{2}},
\end{align}
where
\begin{align}
	f_{1}\left(P_{\mathrm{BS}}^{\mathrm{act}}\right)&=\left(\sigma_{r}^{2} h_{r d}^{2}-\sigma^{2} h_{s r}^{2}\right)\left(P_{\mathrm{BS}}^{\mathrm{act}}\right)^{2}\nonumber\\
	&-2\left(C \sigma_{r}^{2} h_{r d}^{2}+\sigma^{2} \sigma_{r}^{2}\right)  P_{\mathrm{BS}}^{\mathrm{act}} +C \left(C \sigma_{r}^{2} h_{r d}^{2}+\sigma^{2} \sigma_{r}^{2}\right)\!.
\end{align}
If $\sigma_{r}^{2} h_{r d}^{2} = \sigma^{2} h_{s r}^{2}$, $f_{1}\left(P_{\mathrm{BS}}^{\mathrm{act}}\right)$ is a linear function with $f_{1}\left( C/2\right)=0$. Clearly, the optimal solution is  $\left(P_\mathrm{B S}^\mathrm{a c t}\right)^{\star}={C}/{2}$.
When $ \sigma_{r}^{2} h_{r d}^{2}\neq \sigma^{2} h_{s r}^{2}$,  $f_{1}\left(P_{\mathrm{BS}}^{\mathrm{act}}\right)$ is a quadratic function with $f_{1}(0)=C \left(C \sigma_{r}^{2} h_{r d}^{2}+\sigma^{2} \sigma_{r}^{2}\right) > 0$ and $f_{1}(C)=-C \left(C \sigma^{2} h_{s r}^{2}+\sigma^{2} \sigma_{r}^{2}\right) < 0$. Therefore, there must exist one and only one root $\left(P_\mathrm{B S}^\mathrm{a c t}\right)^{\mathrm{rt}}$ for $f_1(P_{\mathrm{BS}}^{\mathrm{act}})$ within $(0,C)$. When $0\leq P_\mathrm{B S}^\mathrm{a c t}<\left(P_\mathrm{B S}^\mathrm{a c t}\right)^\mathrm{rt}$, $f_{1}\left(P_{\mathrm{BS}}^{\mathrm{act}}\right)>0$. When $\left(P_\mathrm{B S}^\mathrm{a c t}\right)^\mathrm{rt}<P_\mathrm{B S}^\mathrm{a c t}\leq C$, $f_{1}\left(P_{\mathrm{BS}}^{\mathrm{act}}\right)<0$. Accordingly, based on (\ref{snr_derivative}), $\gamma_{\text {act }}$ is maximized at $\left(P_\mathrm{B S}^\mathrm{a c t}\right)^{\star} = \left(P_\mathrm{B S}^\mathrm{a c t}\right)^{\mathrm{rt}}$. We next derive the root $\left(P_\mathrm{B S}^\mathrm{a c t}\right)^{\mathrm{rt}}$ which locates in $(0,C)$. After some simplifications, two roots of $f_{1}\left(P_{\mathrm{BS}}^{\mathrm{act}}\right)$ are given by
\begin{align}
	\begin{aligned}
	\left(P_{\mathrm{BS}}^{\mathrm{act}}\right)^\mathrm{RTs}&=\frac{1}{\sigma_{r}^{2} h_{r d}^{2}-\sigma^{2} h_{s r}^{2}} \times \Big\{C \sigma_{r}^{2} h_{r d}^{2}+\sigma^{2} \sigma_{r}^{2} \\
	&\pm\! \sqrt{\left(C \sigma^{2} h_{s r}^{2}+\sigma^{2} \sigma_{r}^{2}\right)\left(C \sigma_{r}^{2} h_{r d}^{2}+\sigma^{2} \sigma_{r}^{2}\right)}\,\Big\}.
\end{aligned}
\end{align}

If $\sigma_{r}^{2} h_{r d}^{2}>\sigma^{2} h_{s r}^{2}$, we have $ C \sigma_{r}^{2} h_{r d}^{2}+\sigma^{2} \sigma_{r}^{2}
=\sqrt{(C \sigma_{r}^{2} h_{r d}^{2}+\sigma^{2} \sigma_{r}^{2})(C \sigma_{r}^{2} h_{r d}^{2}+\sigma^{2} \sigma_{r}^{2})}
>\sqrt{\left(C \sigma^{2} h_{s r}^{2}+\sigma^{2} \sigma_{r}^{2}\right)  \left(C \sigma_{r}^{2} h_{r d}^{2}+\sigma^{2} \sigma_{r}^{2}\right)  } $. Thus, both two roots are positive, and the  left root is the smaller one located in $(0,C)$, as given in (\ref{optimal_power_allocation}). If $\sigma_{r}^{2} h_{r d}^{2}<\sigma^{2} h_{s r}^{2}$, we have $C \sigma_{r}^{2} h_{r d}^{2}+\sigma^{2} \sigma_{r}^{2}<\sqrt{\left(C \sigma^{2} h_{s r}^{2}+\sigma^{2} \sigma_{r}^{2}\right)\left(C \sigma_{r}^{2} h_{r d}^{2}+\sigma^{2} \sigma_{r}^{2}\right)}$. In this case, the left root is negative but the right root is positive. Hence, the right root,  written in (\ref{optimal_power_allocation}), is the solution.

\section{}\label{App_2}
The first-order derivative of $ \left(P_{\mathrm{BS}}^{\text {act }}\right)^{\star} $ with respect to $C$ is
\begin{align}\label{function_appB}
\frac{\partial\left(P_{\mathrm{BS}}^{\text {act }}\right)^{\star}}{\partial C}=\frac{\sigma_{r}^{2} h_{r d}^{2}-\sigma^{2} \sigma_{r}^{2} \sqrt{f_{2}(C)}}{\sigma_{r}^{2} h_{r d}^{2}-\sigma^{2} h_{s r}^{2}},
\end{align}
where $f_{2}(C)=\frac{\left(2 C h_{s r}^{2} h_{r d}^{2}+\sigma^{2} h_{s r}^{2}+\sigma_{r}^{2} h_{r d}^{2}\right)^{2}}{4\left(C \sigma^{2} h_{s r}^{2}+\sigma^{2} \sigma_{r}^{2}\right)\left(C \sigma_{r}^{2} h_{r d}^{2}+\sigma^{2} \sigma_{r}^{2}\right)}$ and
\begin{align}
	&\frac{\partial f_{2}(C)}{\partial C}=\frac{4\left(2 C h_{s r}^{2} h_{r d}^{2}+\sigma^{2} h_{s r}^{2}+\sigma_{r}^{2} h_{r d}^{2}\right) \sigma^{2} \sigma_{r}^{2}}{\left\{4\left(C \sigma^{2} h_{s r}^{2}+\sigma^{2} \sigma_{r}^{2}\right)\left(C \sigma_{r}^{2} h_{r d}^{2}+\sigma^{2} \sigma_{r}^{2}\right)\right\}^{2}} \times \nonumber\\
	&\left\{2 \sigma^{2} \sigma_{r}^{2} h_{s r}^{2} h_{r d}^{2}-\left(\left(\sigma^{2} h_{s r}^{2}\right)^{2}+\left(\sigma_{r}^{2} h_{r d}^{2}\right)^{2}\right)\right\}\overset{(b)}< 0,
\end{align}
where $(b)$ utilizes the inequality $x^2+y^2\geq 2xy$ and $\sigma_{r}^{2} h_{rd}^{2}\neq \sigma^{2} h_{sr}^{2}$ as assumed before. Therefore, the function in (\ref{function_appB}) monotonously increases (decreases) with $C$ if $\sigma_{r}^{2} h_{r d}^{2}>\sigma^{2} h_{s r}^{2}$ ($\sigma_{r}^{2} h_{r d}^{2}<\sigma^{2} h_{s r}^{2}$). Besides, we have $ \lim\limits_{C \rightarrow 0}  \frac{\partial(P_{\mathrm{BS}}^{\text {act }})^{\star}}{\partial C} \rightarrow \frac{1}{2}>0$ and $ \lim\limits_{C \rightarrow \infty}   \frac{\partial(P_{\mathrm{BS}}^{\text {act }})^{\star} }{\partial C} \rightarrow \frac{\sigma_{r}^{2} h_{r d}^{2}-\sqrt{\sigma^{2} \sigma_{r}^{2} h_{s r}^{2} h_{r d}^{2}}}{\sigma_{r}^{2} h_{r d}^{2}-\sigma^{2} h_{s r}^{2}}>0$.
Due to the monotonicity, there must be $\frac{\partial(P_{\mathrm{BS}}^{\text {act }})^{\star}}{\partial C}>0, \forall C>0$. 
Similarly, using $(P_\mathrm{RIS}^\mathrm{a c t})^{\star}=C-(P_\mathrm{B S}^\mathrm{a c t})^{\star}$, we can prove $\frac{\partial(P_{\mathrm{RIS}}^{\text {act }})^{\star}}{\partial C}>0$. Since $C$ increases with $Q_\mathrm{tot}$ but decreases with $N$, the proof is completed.

\section{}\label{App_3}
Define $  \left(P_{\mathrm{BS}}^{\mathrm{act}}\right)^{\star}-\left(P_{\mathrm{RIS}}^{\mathrm{act}}\right)^{\star} \triangleq f_{3}(C) $. Then, we have
\begin{align}
\frac{\partial f_{3}(C)}{\partial C}=\frac{\sigma_{r}^{2} h_{r d}^{2}+\sigma^{2} h_{s r}^{2}-2 \sigma^{2} \sigma_{r}^{2} \sqrt{f_{2}(C)}}{\sigma_{r}^{2} h_{r d}^{2}-\sigma^{2} h_{s r}^{2}},
\end{align}
and $\lim\limits_{C \rightarrow 0}   \frac{\partial f_{3}(C)}{\partial C} \to0 $. Note that we have proved that $f_2(C)$ is a decreasing function. When $\sigma_{r}^{2} h_{r d}^{2}>\sigma^{2} h_{s r}^{2}$, $ \frac{\partial f_{3}(C)}{\partial C} $ is an increasing function leading to $ \frac{\partial f_{3}(C)}{\partial C} >0$ for $C>0$. When $\sigma_{r}^{2} h_{r d}^{2}<\sigma^{2} h_{s r}^{2}$, $ \frac{\partial f_{3}(C)}{\partial C} $ is a decreasing function leading to $ \frac{\partial f_{3}(C)}{\partial C} <0$ for $C>0$.

\section{}\label{App_4}
Using $P_{\mathrm{BS}}^{\mathrm{pas}} = C + N P_{\mathrm{DC}}$, we have $ \frac{\partial}{\partial C} \frac{\left(P_{\mathrm{BS}}^{\mathrm{act}}\right)^{\star}}{P_{\mathrm{BS}}^{\mathrm{pas}}}=\frac{f_{4}(C)}{\left(C+N P_{\mathrm{DC}}\right)^{2}} $ where $ f_{4}(C)=\left(C+N P_{\mathrm{DC}}\right)\left\{\frac{\partial}{\partial C}\left(P_{\mathrm{BS}}^{\mathrm{act}}\right)^{\star}\right\}-\left(P_{\mathrm{BS}}^{\mathrm{act}}\right)^{\star} $ with $f_4(0)=\frac{N P_{\mathrm{DC}}}{2}$. Besides, we have $\frac{\partial}{\partial C} f_{4}(C)=\left(C+N P_{\mathrm{DC}}\right)   \left\{\frac{\partial^{2}}{\partial C^{2}}(P_{\mathrm{BS}}^{\mathrm{act}})^{\star}\right\}$. As proved in Appendix \ref{App_2}, $ \frac{\partial^{2}}{\partial C^{2}}\left(P_{\mathrm{BS}}^{\mathrm{act}}\right)^{\star}>0 $ if  $\sigma_{r}^{2} h_{r d}^{2}>\sigma^{2} h_{s r}^{2}$, which leads to $f_4(C)>f_4(0)>0$ and $  \frac{\partial}{\partial C} \frac{\left(P_{\mathrm{BS}}^{\mathrm{act}}\right)^{\star}}{P_{\mathrm{BS}}^{\mathrm{pas}}}>0 $. By contrast, if $\sigma_{r}^{2} h_{r d}^{2}<\sigma^{2} h_{s r}^{2}$, we have $ \frac{\partial^{2}}{\partial C^{2}}\left(P_{\mathrm{BS}}^{\mathrm{act}}\right)^{\star}<0 $ and then $f_4(C)$ is a decreasing function. As $C\to\infty$, we have 
\begin{align}
f_{4}(\!\infty\!)\! \rightarrow \!\!
\frac{\sigma^{2} \sigma_{r}^{2} (\frac{\sigma^{2} h_{s r}^{2}+\sigma_{r}^{2} h_{r d}^{2}}{2 \sqrt{  \!  \sigma^{2} h_{s r}^{2} \sigma_{r}^{2} h_{r d}^{2}}} \!-\!1  \!  ) \!}{\sigma_{r}^{2} h_{r d}^{2}-\sigma^{2} h_{s r}^{2}}
\!+\!
N \! P_{\mathrm{DC}} \frac{\sigma_{r}^{2} h_{r d}^{2} \!-\! \!  \sqrt{ \!\!  \sigma^{2} \sigma_{r}^{2} h_{s r}^{2} h_{r d}^{2}}}{\sigma_{r}^{2} h_{r d}^{2}-\sigma^{2} h_{s r}^{2}} \! , \!
\end{align}
where the first term is negative while the second term is positive. Thus, when $N$ is larger than a threshold, $f_4(C)>f_{4}(\infty)>0$  and then $  \frac{\partial}{\partial C} \frac{\left(P_{\mathrm{BS}}^{\mathrm{act}}\right)^{\star}}{P_{\mathrm{BS}}^{\mathrm{pas}}}>0 $. Otherwise, $f_4(C)$ decreases from positive value $f_4(0)$ to negative value $f_4(\infty)$ which means that $ \frac{\left(P_{\mathrm{BS}}^{\mathrm{act}}\right)^{\star}}{P_{\mathrm{BS}}^{\mathrm{pas}}} $ firstly increases but then decreases.

\end{appendices}
\bibliographystyle{IEEEtran}
\bibliography{myref.bib}
\end{document}